# Superconducting Properties of MgB$_2$ Bulk Materials Prepared by High Pressure Sintering

Y. Takano, H. Takeya, H. Fujii , H. Kumakura, T. Hatano and K. Togano

*National Research Institute for Metals, 1-2-1, Sengen, Tsukuba 305-0047, Japan.*
*CREST, Japan Science and Technology Corporation, 2-1-6, Sengen, Tsukuba 305-0047, Japan*

H. Kito and H. Ihara

*Electrotechnical Laboratory, 1-1-4, Umezono, Tsukuba 305-8568, Japan*

High-density bulk materials of a newly discovered 40K intermetallic MgB$_2$ superconductor were prepared by high pressure sintering. Superconducting transition with the onset temperature of 39K was confirmed by both magnetic and resistive measurements. Magnetization versus field (M-H) curve shows the behavior of a typical Type II superconductor and the lower critical field H$_{c1}$(0) estimated from M-H curve is 0.032T. The bulk sample shows good connection between grains and critical current density J$_c$ estimated from the magnetization hysteresis using sample size was 2x10$^4$A/cm$^2$ at 20K and 1T. Upper critical field H$_{c2}$(0) determined by extrapolating the onset of resistive transition and assuming a dirty limit is 18T

Very recently, J. Akimitsu[1] has announced the discovery of 39K superconductivity in MgB$_2$, which has stimulated considerable interest in it as a new family of high temperature superconductors. The transition temperature T$_c$ is extremely higher than those previously reported for Nb$_3$Ge(T$_c$=23.2K)[2] and YPd$_2$B$_2$C(T$_c$=23K)[3] as the highest T$_c$ values in A15 intermetallic compounds and intermetallic borocarbides, respectively. The discovery seems to confirm the speculation that the low mass of boron should be conducive to high phonon frequencies and consequently high temperature superconductivity.[4] Not only due to this basic interest but also in order to prove the possibility of this new material for various practical applications, the accumulation of experimental data on various physical properties related to the superconductivity is an urgent need.

Since Mg is extremely volatile at elevated temperatures, it is very difficult to prepare a bulk material with high-density sufficient for transport measurements by sintering at an ambient pressure. In this paper, we report the preparation of high-density MgB$_2$ bulk samples using high pressure sintering. Using those bulk samples, we performed magnetization and transport measurements in applied magnetic fields in order to evaluate various electromagnetic properties such as lower critical field H$_{c1}$(0), upper critical field H$_{c2}$(0) and critical current density J$_c$ of the MgB$_2$ superconductor.

Starting material used in this study was the powder of MgB$_2$ (99%, -100Mesh, Furuuchi Chemical), which is commercially available as a reagent. Our x-ray diffraction experiment shows that although the major phase of the powder is MgB$_2$ with AlB$_2$ structure, small peaks of MgO as an impurity phase were observed in the diffraction pattern. Bulk materials were prepared by high pressure sintering using a cubic anvil press (RIKEN CAP-07). The powder was encased in a BN crucible and subjected to a high pressure of 3.5GPa. During the application of high pressure, the samples were heated at 775°C, 1000°C, and 1250°C, respectively, for 2h by a carbon heater which surrounded the BN crucible. The sintered samples were taken out by breaking the BN crucible and the surface layer of about 0.3mm thickness was ground off. The samples finally obtained have a cylinder shape whose size is about 4mm in diameter and 3.5mm in height.

The cross sections of the samples were polished and observed by optical microscopy (OM) and scanning electron microscopy (SEM). The as-polished sections are metallic and no porosity is observed, indicating that the samples have an extremely high density. The density measured was 2.66g/cm$^2$, which almost corresponds to the calculated value of 2.63g/cm$^2$ using lattice parameters. The samples are very hard and the Vickers hardness is Hv=1700-2800. Figure 1 shows the microstructure of the 775°C sintered sample observed by SEM, which is composed of a dark matrix and white particles. The powder x-ray diffraction pattern of the 775°C sample is almost same as that of the starting MgB$_2$ powder. This indicates that the matrix in Fig. 1 is MgB$_2$ phase and the white particles are MgO, which was present in the starting MgB$_2$ powder as an impurity phase. The volume fraction of MgO estimated from the figure is about 10%. The 1000°C sintered sample shows a similar microstructure, however, the volume fraction of MgO estimated from the SEM photograph is increased to about 15%. In addition, the powder x-ray diffraction pattern of the 1000°C sample shows the existence of a small amount of MgB$_4$ as an impurity phase, indicating the partial decomposition of MgB$_2$ to MgB$_4$ at this sintering temperature. The 1250°C sample shows an apparently inhomogeneous structure in macroscopic scale on the cross section, the color changing from the circumference to the center, indicating the occurrence of contamination from the crucible wall. Therefore the 1250°C sample was excluded from the superconductivity measurement.

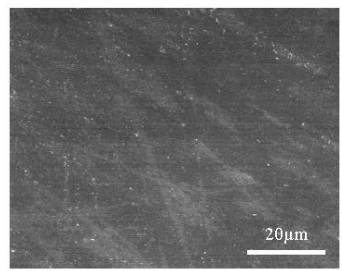

FIG. 1. Scanning electron micrograph of the MgB$_2$ bulk sample sintered at 775°C under high pressure.

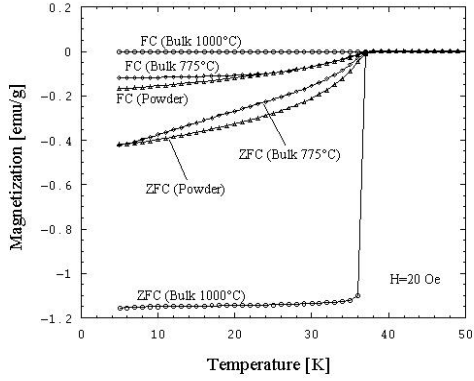

FIG. 2. Temperature dependent magnetization curves for the MgB$_2$ starting powder and the bulk samples sintered at 775°C and 1000°C under high pressure. FC and ZFC are the field cooling and zero field cooling curves, respectively.

Temperature dependence of magnetization was measured at 20Oe by a SQUID magnetometer. Figure 2 is the results for the starting MgB$_2$ powder and the bulk samples sintered at 775°C and 1000°C. The starting powder has a relatively broad superconducting transition with the onset temperature of 39K. It is not clear what causes this broad transition, however, it may be due to the stress induced by powdering process. The 775°C sample has also a relatively broad superconducting transition, indicating that the sintering temperature of 775°C is too low to improve the crystallinity. However, the superconducting transition was drastically improved for the 1000°C sample, showing a very sharp transition. The large difference between the zero-field cooling and field cooling curves of the 1000°C sample indicates that the material has a fairly large flux pinning force resulting in the trapping of magnetic flux in the field cooling condition.

Resistive measurement was also performed for the bulk samples. Figure 3 shows the resistive superconducting transitions under magnetic fields up to 9T. The normal state resistivity showed a large difference between the 775°C and 1000°C samples, 360mΩ·cm and 60μΩ·cm, respectively, at the onset of transition. This indicates that the 1000°C sample has much better grain connectivity. The figure shows the shift of transition to lower temperatures with increase of the magnetic field. Even onset temperature decreases with the increasing field, which is in contrast with the high-Tc cuprate superconductors and similar to A15 intermetallic compounds such as Nb$_3$Sn. In accordance with the magnetization data shown in Fig. 2, the 775°C sample shows a relatively broad transition, while the 1000°C sample has a sharp transition with the onset temperature of 39 K and the transition width of 1K in zero applied magnetic field. In applied magnetic fields, the superconducting transition becomes broader, however, the broadening is much smaller for the 1000°C sample. From these results, we speculate that the broadening is not caused by the intrinsic nature such as anisotropic electronic state that is a striking nature of high-Tc cuprate superconductors but by the weak link between the grains. Although exact determination of the superconducting parameters requires the experiments using high quality single crystal, we tried to estimate the upper critical field, $H_{c2}$, using the resistive onset temperature of the 1000 °C sample. Figure 4 shows the plot of onset temperature for applied magnetic fields up to 9T. The curve shows a positive curvature very near to $T_c$ similar to the borocarbide systems.[5] Except this region, the curve is linear, whose gradient $-dH_{c2}/dT$ is 0.74T/K. Therefore, linear extrapolation gives the $H_{c2}$ value of 26.5T. For the type II superconductor in the dirty limit, $H_{c2}(0)$ is given by the formula, $H_{c2}(0)=0.691 \times dH_{c2}/dT \times T_c$.[6,7] The $H_{c2}(0)$ value of $MgB_2$ obtained by this equation is 18T.

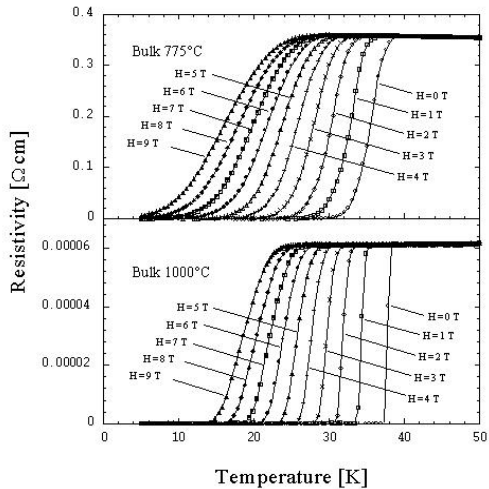

FIG. 3. Resistive superconducting transitions for the bulk sample sintered at 775°C and 1000°C under high pressure.

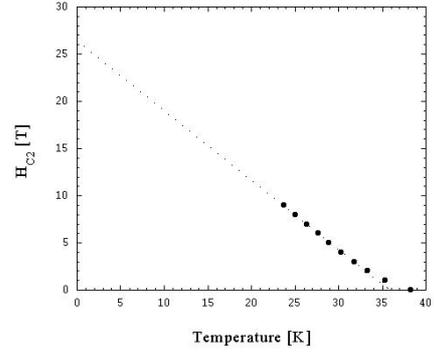

FIG. 4. $H_{C2}$(T) plot defined by the onset temperature of resistive transition measured in magnetic field for the 1000°C sample.

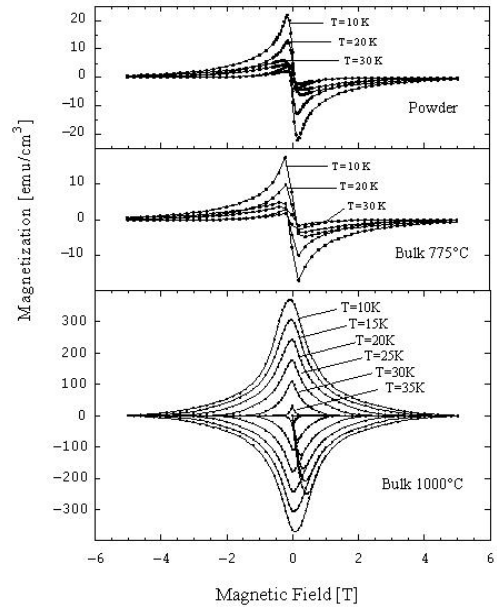

FIG. 5. Magnetization versus magnetic field (M-H) curves for the MgB$_2$ starting powder and the bulk samples sintered at 775°C and 1000°C under high pressure.

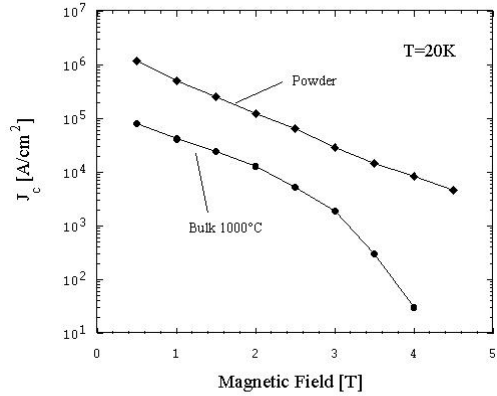

FIG. 6. $J_c$'s at 20K as a function of applied magnetic field for the $MgB_2$ powder and bulk sample sintered at 1000°C under high pressure.

Using a SQUID magnetometer, magnetization versus magnetic field (M-H) curves were measured at several temperatures up to 5T. Figure 5 shows the results of the starting $MgB_2$ powder and the 775°C and 1000°C samples. The figure shows the characteristic curve of type II superconductors. The magnetization hysteresis of the 1000°C sample is much larger than that of the 775°C sample, indicating that the $MgB_2$ grain connectivity is much improved by the 1000°C sintering. From hysteresis of the magnetization curve, $\Delta M(emu/cm^3)$, we can estimate critical current density $J_c$ on the assumption of a critical-state model with the simple formula, $J_c= 30 \times \Delta M /d$, where d is the size of the sample(cm).[8] Figure 6 shows $J_c$'s at 20K as a function of applied magnetic fields calculated for the $MgB_2$ powder and 1000°C sample. The calculation was done by using the average particle size for the $MgB_2$ powder, which is 1μm measured by SEM, and the sample size for the 1000°C bulk sample. The $J_c$ value at 20K and 1T for the $MgB_2$ powder is $5 \times 10^5 A/cm^2$, which shows the $MgB_2$ has a fairly large pinning force in the grain. However, the $J_c$ value of the 1000°C bulk sample at 20K and 1T decreases to $4 \times 10^4 A/cm^2$ indicating that the sample still has a weak link problem. From M-H curves, we also tried to determine the lower critical field, $H_{c1}(0)$. The $H_{c1}(T)$ was defined as the magnetic field, where the initial slope meets the extrapolation curve of $(M_{up}+M_{down})/2$ and plotted as a function of the temperature for the $MgB_2$ powder as shown in Fig. 7. Extrapolation of the plot gives the $H_{c1}(0)$ value of 0.032T.

In summary, in order to measure the transport properties of a newly discovered $MgB_2$ superconductor, we have prepared high-density bulk materials using high pressure sintering. High temperature superconductivity with the onset temperature of 39K was confirmed by both magnetic and resistive measurements. The bulk sample sintered at 1000°C showed a good electrical connection between the grains and the critical current density $J_c$ estimated from the magnetization hysteresis using the sample size was $2 \times 10^4 A/cm^2$ at 20K and 1T. The M-H curve shows the typical behavior of type-II superconductor. The $H_{c1}(0)$ extrapolated from $H_{c1}(T)$ values determined from the M-H curves is 0.032T. The determination of $H_{c2}(0)$ was also attempted using the onset temperatures of resistive transitions in the fields. Assuming the type II superconductivity in dirty limit, the $H_{c2}(0)$ value of 18T was obtained by the extrapolation. However, since the $MgB_2$ is expected to have a two-dimensional electronic state,[9] the exact determination of the superconducting parameters will require single crystals.

We would like to thank to Prof. M. Tachiki, Dr. S. Arisawa and Dr. R.Jayavel of National Research Institute for Metals, Tsukuba, Japan for useful discussions.

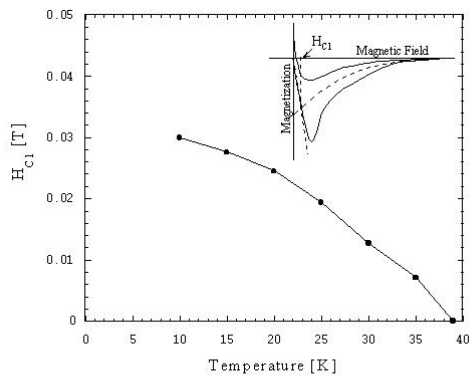

FIG. 7. $H_{c1}(T)$ plot determined using the M-H curves of the $MgB_2$ powder sample.